\documentclass[journal]{new-aiaa}
\usepackage[utf8]{inputenc}
\usepackage{textcomp}
\usepackage{makecell}
\usepackage{array}
\usepackage{graphicx}
\graphicspath{ {./images/} }
\usepackage{amsmath}
\usepackage[version=4]{mhchem}
\usepackage{siunitx}
\usepackage{float}
\usepackage{longtable,tabularx}
\usepackage{xcolor}
\usepackage{booktabs}
\usepackage{soul}
\usepackage{biblatex}
\pagebreak
\begin{document}
\hspace{0pt}
\vfill

\title{Machine Learning Interventions for Weed Detection using Multispectral Imagery and Unmanned Aerial Vehicles - A Systematic Review}

\author{Drishti Goel \footnote{Research Fellow, Microsoft, Bengaluru, India} and Bhavya Kapur\footnote{Data Scientist, NeenOpal Intelligent Solutions Inc., Bengaluru, India} and Prem Prakash Vuppuluri\footnote{Assistant Professor, Dayalbagh Educational Institute (Deemed University), Agra, India}}

\maketitle
\vfill
\hspace{0pt}
\pagebreak

\begin{center}
    \textbf{Abstract}
\end{center}

The growth of weeds poses a significant challenge to agricultural productivity, necessitating efficient and accurate weed detection and management strategies. The combination of multispectral imaging and drone technology has emerged as a promising approach to tackle this issue, enabling rapid and cost-effective monitoring of large agricultural fields. This systematic review surveys and evaluates the state-of-the-art in machine learning interventions for weed detection that utilise multispectral images captured by unmanned aerial vehicles. The study describes the various models used for training, features extracted from multi-spectral data, their efficiency and effect on the results, the performance analysis parameters, and also the current challenges faced by researchers in this domain. The review was conducted in accordance with the PRISMA guidelines. Three sources were used to obtain the relevant material and the screening and data extraction were done on the COVIDENCE platform. The search string resulted in 600 papers from all sources. The review also provides insights into potential research directions and opportunities for further advancements in the field. These insights would serve as a valuable guide for researchers, agricultural scientists, and practitioners in developing precise and sustainable weed management strategies to enhance agricultural productivity and minimise ecological impact.


\section{Introduction}

Weed infestation and monitoring is one of the most critical and expensive problems in agriculture \cite{PIMENTEL2005273, RAO2020104451}. Weeds can utilise up to 30\% to 40\% of the soil micronutrients, moisture, and other essential elements, which results in depleted resources for the dominant crops \cite{Woyessa_2022}. This results in low crop yield, efficiency, quality, and increased labor and machinery costs. This continuous competition with the crops has a colossal negative effect on the productivity of the farmers, specifically in developing countries. Moreover, weeds also act as a refuge for other elements of the environment that are harmful to the crops, such as pests, bacteria, fungi, etc \cite{article3, article4, Roshan2019}.

Conventional methods of weed detection and removal include manual removal and blanket spraying of chemical herbicides across the entire field. Manual removal is expensive, time-consuming, and practically inefficient due to the general scarcity of workers and funds. Moreover, blanket spraying of herbicides has an adverse effect on the health of the crop, and with the current shift towards organic and minimalistic farming, various researchers and even the United Nations' Food and Agriculture Organization (FAO) are encouraging research in the development of novel and reliable mechanisms for weed detection and removal \cite{online3}. 

The European Green Deal \cite{c20, c20a}, aims to make Europe the world's first climate-neutral continent by the year 2050. They highlight the importance of a safe, sustainable, and reliable food system that minimises the utilization of fertilisers, pesticides, and antimicrobials, which simultaneously increase organic farming, improve animal welfare, and reverse biodiversity degradation. They propose a "Farm to Fork Strategy" that lays an approach to achieve the goals to build a food chain that works for the consumers, producers, climate, and the environment. 

Weed infestations cause a 31.5\% decrease in plant production across the globe, resulting in an annual economic loss of approximately USD 32 billion \cite{agronomy12081808}. Conventional methods of weed removal are not only practically inefficient, expensive, and hazardous for consumer health, but methods such as blanket spraying of herbicides have proven to become ineffective over time, due to the development of immune mechanisms in the weed infestations, because of the overuse of such chemicals.

All these factors have led to research in developing and integrating technologies such as the Internet of Things, Unmanned Aerial Vehicles (UAVs), and Machine Learning, in order to leverage the potential they possess of identifying and distinguishing weeds from crops, accurately, and economically \cite{su13094883, REJEB2022107017}. 

Machine Learning is a technique that can be used to train reliable systems to make intelligent decisions or perform tasks without explicit programming using a set of training patterns and features. Machine Learning has been a key contributor to the advancement of "Smart Agriculture", a field that uses such technologies to make decisions related to crop plantation, irrigation, herbicide and pesticide application, and weed detection \cite{s22166299}. Various models have been employed to tackle this problem, and techniques such as supervised models, unsupervised models, neural networks, and deep learning have been tested in several different situations, and have shown great potential to solve this problem. 
These models are usually trained on images of agricultural fields/crops. A prominent method of capturing these images is by using UAVs. UAVs provide an efficient method of capturing continuous feed (images or video) of the field under study. They provide information on the fields from a bird's-eye view and the routes of the UAVs can be pre-defined, hence, requiring limited human assistance in image acquisition. The benefits of UAVs, their application to obtaining data, and metrics that could be used to assess crop health, and differentiate between vegetation-soil or vegetation-weed for efficient management are described by Rango et. al. \cite{Rango}.

RGB or multispectral camera sensors are mounted on the UAVs, which capture images of the field. RGB sensors capture spatial information of the fields by capturing three (Red, Green, and Blue) low-resolution wavebands. These cameras are most prominently used because they are low-cost, economical, accessible, and convenient to use, with simple data processing and minimal requirements \cite{Yang2017UnmannedAV}. They can efficiently be deployed in a variety of weather conditions, with simple modifications, however, they fail to capture the non-visible light of the electromagnetic spectrum, making it relatively difficult to distinguish between objects that appear similar to the human eye. Multispectral and hyperspectral cameras, on the other hand, overcome this problem by capturing light from the visible and invisible spectrum with high working efficiency and fast frame imaging \cite{Yang2017UnmannedAV}. They capture the reflectance and absorption characteristics of the crops which are used for crop phenotyping. 

In this review, the literature at the intersection of these technologies is surveyed, focusing on the application of machine learning for identifying weed regions in an agricultural field with the help of UAV-acquired multispectral imagery.

\section{Background and Related Work}

This structured study focuses on addressing a set of predefined questions by conducting a detailed and comprehensive search strategy across the entire literary search space. It differs from traditional reviews, as the latter are often subject to a selection bias due to the absence of a systematic search strategy and plan. 
The essence of this review is its methodical search strategy, and meta-analysis, which presents the data and literature in a concise, effective, and targeted manner, without the author’s narrative bias.

To conduct this systematic review successfully, we identified, not only the most effective search strategy but also the parameters to gauge the relevance of the studies for answering the pre-defined search questions. Following the PRISMA statement \cite{MOHERnew, MOHER2010336}, which consists of a list of 27 checklist points and a four-phase flow diagram, we present the report in a standardised manner.

A detailed analysis of the current and future trends in the field of weed detection using UAVs was presented in a study \cite{agriculture11101004}, that reported using the PRISMA guidelines, and conducted their search on Scopus, ScienceDirect, Commonwealth Agricultural Bureaux (CAB) Direct, and Web of Science databases between the 1st of January 2016 and the 18th of June 2021. They highlighted the prevalent UAV detection applications, types of cameras, their specification, and the advantages and disadvantages of the different classification algorithms that were used in the studies.

Another study compared the performance of various machine learning algorithms such as SVM, ANN, and CNN to distinguish between crops and weeds using spectroscopy, color imaging, and hyperspectral imaging. It suggested that hyperspectral images successfully capture the spectral and image features of plants, and highlight the growth in the number of studies that employ multispectral imaging, being a more resource-efficient reformation of hyperspectral imaging techniques \cite{app12052570, article1}.

Considering the rapidly increasing interest in using multispectral sensors for weed detection using Machine Learning and UAVs, and its potential advantages as presented in the previous research works, this study aims to investigate the work done and the results obtained by using these technologies; underlining the currently existing practices.

\section{Methodology}
\subsection{\emph{Research Questions}}
\textit{RQ1:} What are the most prevalently used UAV and camera models? 

\textit{RQ2:} What features are extracted from the captured data?

\textit{RQ3:} Which machine learning algorithms have been applied for weed detection?

\textit{RQ4:} Which are the evaluation approaches and parameters used?

\textit{RQ5:} What are the major challenges in this domain?

\subsection{\emph{Search Strategy}}
To obtain the relevant studies for this review, a search strategy was formed iteratively. Three databases were used, as shown in Table \ref{tab1}, which were chosen following the review conducted by Abderahman Rejeb et al. \cite{agronomy11071435}, which presented a list of the most commonly used databases for UAV-related studies.
In this study the search string for each database, as shown in Table \ref{tab2}, was developed and modified as more synonymous terms were identified iteratively.

\begin{table}[hbt!]
\caption{Databases included in the systematic search strategy}
\centering
\begin{tabular}{lcccccc}
\hline
S.No.& & Database Name\\\hline
1& & IEEE Robotics and Automation Letters&\\
2& & MDPI (Sensors, Agronomy, Remote Sensing)&\\
3& & Web Of Science&\\
\hline
\label{tab1}
\end{tabular}
\end{table}
\begin{table}[hbt!]
\caption{Search strings used for each database}
\begin{tabular}{@{}llc@{}}
\toprule
\textbf{Database Name}                   & \textbf{No. of Results} & \textbf{Search String}                                                                                                                                                                                                                                        \\ \midrule
IEEE Robotics and Automation Letters     & \multicolumn{1}{c}{37}  & \begin{tabular}[c]{@{}c@{}}(Agriculture OR Farming OR Agricultural OR Farmers) \\ AND (UAV OR Unmanned Aerial Vehicle OR Drones) \\ AND (Machine Learning OR Artificial Intelligence OR ML) \\ AND (Weeds OR Weed OR Weed Removal OR De-weeding)\end{tabular} \\
                                         &                         & \multicolumn{1}{l}{}                                                                                                                                                                                                                                          \\
MDPI (Sensors, Agronomy, Remote Sensing) & \multicolumn{1}{c}{503} & \begin{tabular}[c]{@{}c@{}}(Agriculture OR Farming OR Agricultural OR Farmers) \\ AND (UAV OR Unmanned Aerial Vehicle OR Drones) \\ AND (Machine Learning OR Artificial Intelligence OR ML) \\ AND (Weeds OR Weed OR Weed Removal OR De-weeding)\end{tabular} \\
                                         &                         & \multicolumn{1}{l}{}                                                                                                                                                                                                                                          \\
Web Of Science                           & \multicolumn{1}{c}{7}   & \begin{tabular}[c]{@{}c@{}}ALL=((Agriculture OR Farming ) \\ AND ( unmanned aerial vehicle OR UAV OR Drones) \\ AND (Machine Learning OR Artificial Intelligence OR ml) \\ AND (Weeds or de-weeding))\end{tabular}                                            \\ \bottomrule
\label{tab2}
\end{tabular}
\end{table}

\subsection{\emph{Selection Criteria}}
To ensure that the data we collect is relevant to the questions that are to be addressed through this review, the following inclusion and exclusion criteria were established: 

\subsubsection{\textit{Exclusion Criteria-}}
a. Papers that were unavailable, even after a full-text request 

b. Papers that had not been published

c. Secondary studies, such as reviews and surveys

d. If the study was conducted on terrains other than land-agricultural fields (eg. water body analysis)

\subsubsection{\textit{Inclusion Criteria-}}

a. Primary studies

b. Studies describing Machine Learning interventions that target weed detection using UAV-acquired multispectral images. 

c. Projects that include work implemented on actual datasets (either acquired/developed by authors themselves or drawn from publicly available data)

d. The data which is collected/used should comprise of multispectral images and not RGB

\subsection{\emph{Studies Selection}}
\begin{figure}[hbt!]
\centering
\includegraphics[width=0.8\textwidth]{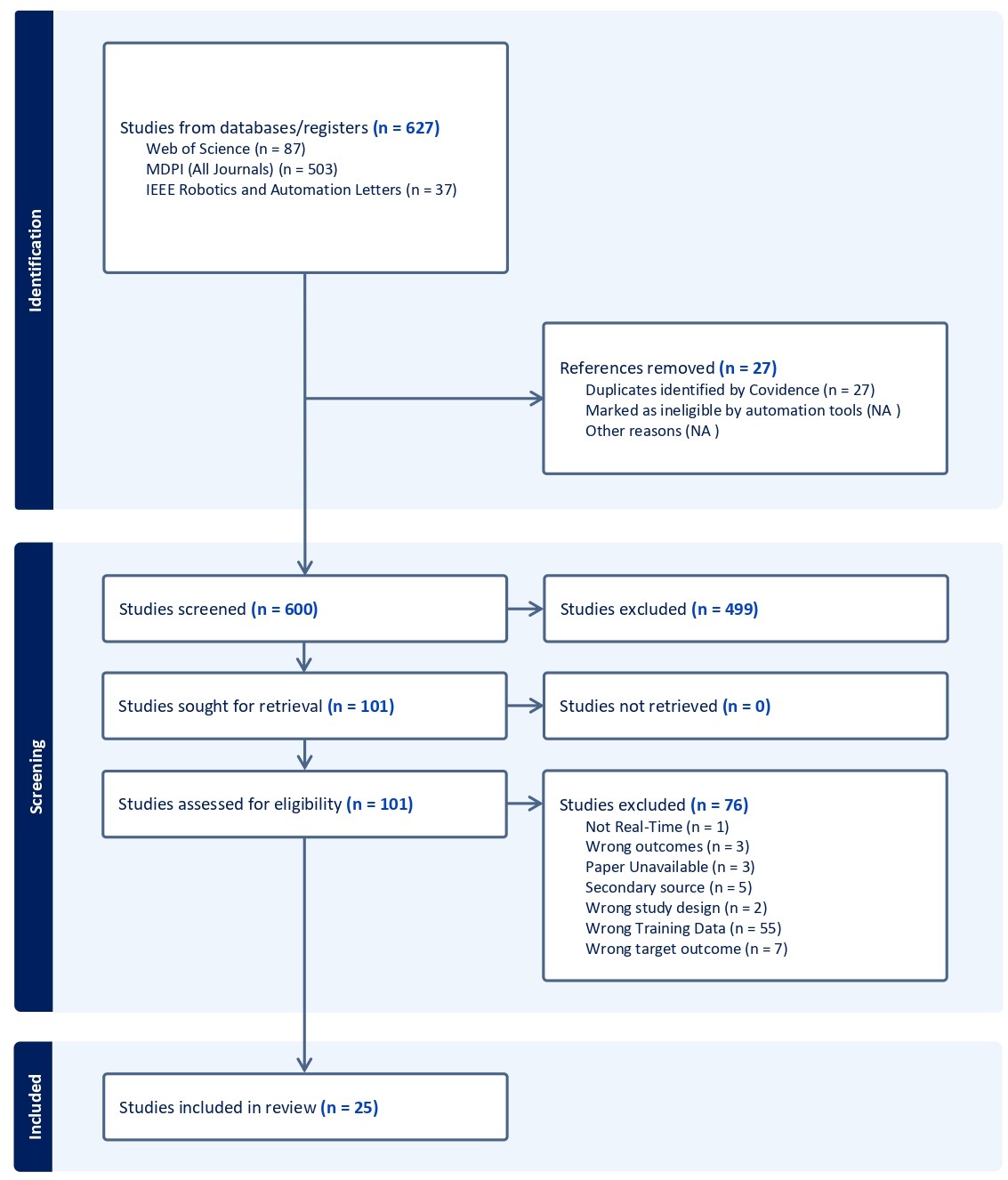}
\caption{Description of the study-selection process using the PRISMA flowchart diagram}
\label{fig1}
\end{figure}

This section details the process that was employed for conducting the review, which is described in Fig. \ref{fig1}. After importing all the papers from the various sources on Covidence, the duplicates were removed automatically. Out of the 627 papers imported initially, 27 duplicates were identified and removed. 

Next, in the Title and Abstract screening, the authors individually evaluated the papers based on the inclusion/exclusion criteria and resolved any conflicts regarding the categorization of the papers together. At this stage, elimination should be done leniently, only when the papers are evidently violating any of the inclusion criteria.  499 studies were removed after the Title and Abstract screening, with mutual agreement among the authors. The remaining 101 studies were taken to the next stage of the Full Text Review, wherein the authors individually categorised the papers as “Include”, “Exclude” or “Maybe” based on the full-text analysis. Here, the papers were carefully analyzed after reading the entire paper, the experimental procedure followed and the results reported. This process is critical and is facilitated with the help of intricately designed inclusion/exclusion criteria. The full-text review resulted in 21 papers, which were finally considered for data extraction and analysis. 

\subsection{\emph{Data Extraction, Synthesis, and Reporting}}
Once the final studies were selected after the full-text review, the next step was to extract the important data from those papers, which would ultimately help in answering the research questions as mentioned in Section V(A).
To this end, the data extracted from each paper consisted of the following details: the name of the paper, the author(s) of the paper, the institution to which the author(s) belongs, the date of publication of the paper, the source from which the paper was obtained, the publication document type, the country in which the study was conducted, what kind of data was used to train the model, the features of the data, any image processing software used to preprocess the data, the crop that the study was conducted on, which weeds were primarily being targeted, the camera used to obtain the images for the dataset and the specifications of that camera, which UAV model and specifications were used, which machine learning algorithm(s) were used to train the model for weed detection, evaluation metrics for the algorithm(s) used, the drawbacks of the methodology used, any specific use-case of the research and other additional notes.
The collected data was then synthesised by the authors to create a comprehensive picture of all the information obtained. This data was further used to answer the research questions put forth in Section V(A).

\subsection{\emph{Preliminary Analysis}}
In this section, the preliminary analysis and their findings are presented. This analysis was carried out using only the final 21 research papers, whose full-text review was done by the authors. The different parameters on which analysis has been carried out are the year of publication, the crop being used in the study and the weeds being detected in the study.

\subsubsection{Publications Based On Year}
\begin{figure}[H]
\centering
\includegraphics[width=.8\textwidth]{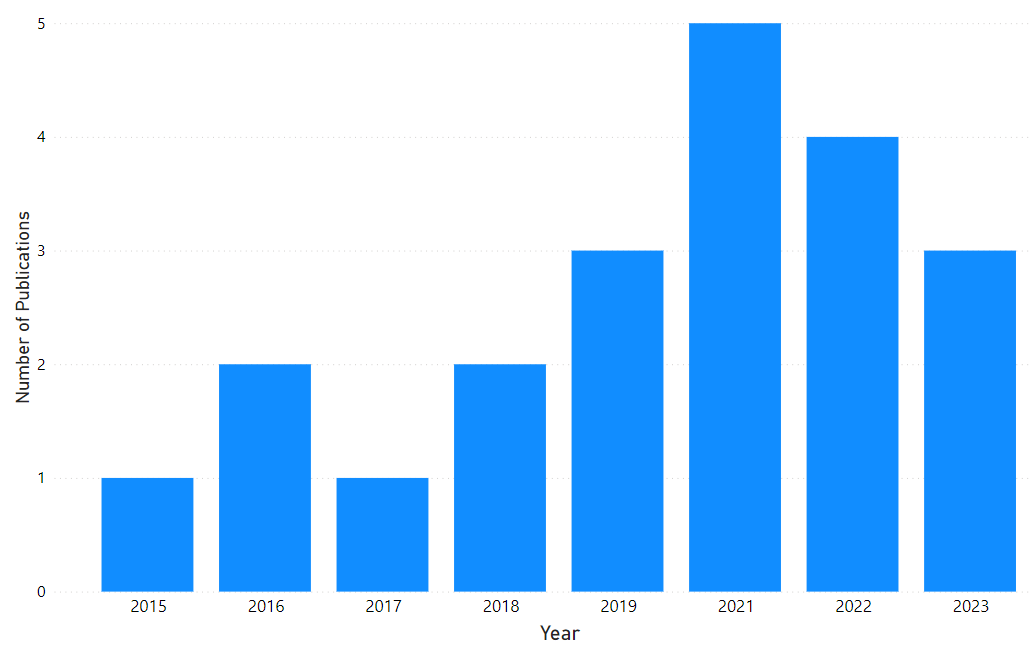}
\caption{Publication trends of the selected studies over the years}
\label{fig2}
\end{figure}

The Fig. \ref{fig2} shows the number of papers chosen per year, from the year 2015 to 2023. According to the results, the maximum number of papers have been published in the year 2021, i.e., 23.8 \% of all the papers have been published in 2021, while the bulk of papers belong to the years 2019, 2021, 2022, and 2023.

\subsubsection{Publications Based On Crops}
The selected research papers report research on a total of 12 crops: Sugar Beet, Sunflower, Wheat, Soybean, Sorghum, Cotton, Coffee, Corn, Sesame, Olive Groves, Maize, and Peanut Seedlings. 

Another category, used in several publications, is ‘Fields with a mix of random crop species’. This includes fields with random vegetation but a particular target weed that is co-dominant along with the vegetation. 

S. marianum weed at Thessaloniki, Greece \cite{s17092007} \cite{10.1080/01431161.2016.1252475} \cite{jimaging4110132}, mouse-ear hawkweed in a floristically simple grassland \cite{rs15061633}, and an abandoned plot of land in Hungary with common Milkweed \cite{land10010029} are examples of this particular category of crops considered in this review.

\begin{figure}[H]
\centering
\includegraphics[width=1.0\textwidth]{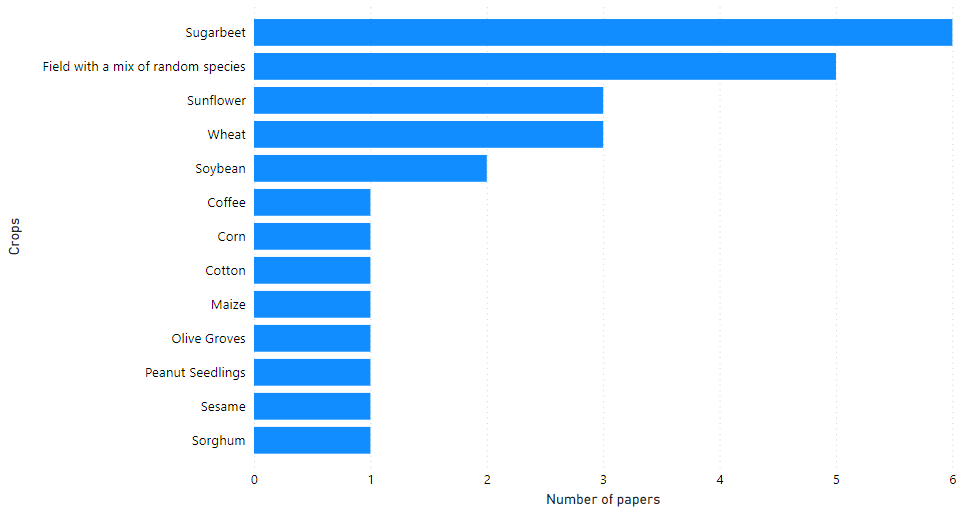}
\caption{Number of selected papers across crop types}
\label{fig3}
\end{figure}

From the results shown in Fig. \ref{fig3}, it is evident that Sugar Beet is the crop most studied in the selected research papers, i.e., it is studied in 6 research papers. It is followed closely by the category ‘Fields with a mix of Random Species’, which is studied in 5 of the papers.

\subsubsection{Publications Based On Weeds}
To properly train a model to classify crops and weeds, it is important to clearly annotate the training data, i.e., clearly mark which portion of the image is the crop, and which is the weed. To do that, it is essential to have an understanding of which weeds are present among the crops, and which physical features can be used to differentiate them from the crop.

A total of 26 weeds were used in the training of models in the selected research papers. These weeds included S. marianum, Amaranth, Pigweed, Black Grass, Hawkweed, Morning Glory, Brachiaria Decumbens, Mallow Weed, LiverSeed Grass, Bellive, Bindweed, Turnsole, Knotweed, Canary Grass, Common Milkweed, Galinsoga, Atriplex, Polygonum, Kochia, Marestail, Common Lambsquarters, Avena Sterilis, Chenopodium album, Humulus scandens, Xanthium Sibiricum and Gramineae.

Another category that was created by the authors of this study is ‘Assorted’, which categorises all the studies wherein multiple weeds were present in the field, and the machine learning algorithm targeted to identify the crop and classify all the other weeds as a single category. This can be seen in some papers in which three learning methods were used to distinguish the crops from the weeds \cite{7849987, PEREZORTIZ2015533}. The central idea of these papers is that by considering the position of weeds in relation to crop rows, it is possible to enhance the ability of weed discrimination. 

\begin{figure}[hbt!]
\centering
\includegraphics[width=.7\textwidth]{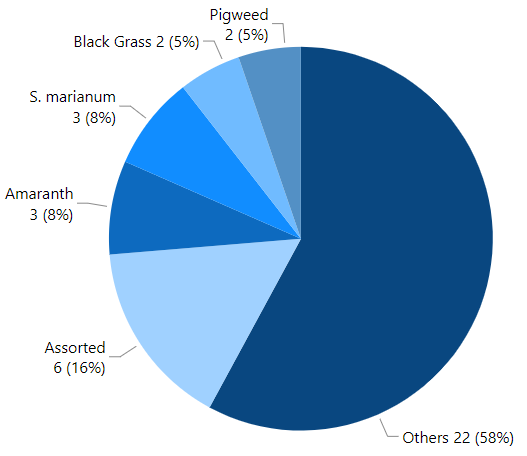}
\caption{Publications based on weeds}
\label{fig4}
\end{figure}

Moreover, in a sesame field, hyperspectral images were obtained by a UAV, and the crop and weed patches were manually labeled and used to train a novel SesameWeedNet convolutional neural network \cite{DENG2018298}. Since no particular weed was specified, this paper is also considered in the ‘assorted weeds' category, as shown in Fig. \ref{fig4}. 

In another study, three weeds, kochia, marestail, and common lambsquarters were captured in a sugar beet field, and a feedforward neural network was trained to classify these weeds against the crop as well as against each other \cite{Scherrer2019HyperspectralIA}. Different biotypes of these weeds were also used, namely, herbicide-susceptible and herbicide-resistant kochia biotypes.

The examination of various weed species within the chosen research papers revealed that the 'assorted weeds' were the most frequently investigated, while S. marianum and Amaranth, emerged as the most extensively researched individual weed species.

S. marianum was usually found in fields with a random mix of crop species \cite{s17092007, 10.1080/01431161.2016.1252475, jimaging4110132}, while Amaranth was found in the crops of sorghum, sugar beet, and soybean \cite{agronomy11071435, rs10091423, agronomy11101909}.

\section{RQ1: What are the most prevalently used UAV and camera models}
In this section, we highlight the most prevalently used UAV devices and the camera models, with their respective specifications.
Nine different UAV models of Dà-Jiāng Innovations (DJI) were used in several studies, followed by the senseFly eBee and MD4-1000 UAVs, each of which was used in three of the selected studies, as shown in Table \ref{tab3}. 

\begin{table}[H]
\caption{UAV models used in the selected research papers}
\begin{tabular}{lcl}
\hline
\textbf{UAV Model}                                                                                                                                                                             & \textbf{Number of Times Used} & \textbf{References}                                                         \\ \hline
\begin{tabular}[c]{@{}l@{}}DJI Inspire 2, DJI Mavic Pro, DJI S1000,\\ DJI Matrice 600, DJI Matrice 600 Pro,\\ DJI Matrice 100, DJI Matrice 210,\\DJI Phantom 3, DJI Phantom 3 Pro\end{tabular} & 9                         & {\cite{rs10091423}}, {\cite{SU2022106621}}, {\cite{rs15061633}}, {\cite{Scherrer2019HyperspectralIA}}, {\cite{agronomy13030830}},{\cite{10.1007/978-3-030-29888-3_45}}, {\cite{DENG2018298}}, {\cite{Wang2022WeedDE}} \\
                                                                                                                                                                                               &                           &                                                                             \\
senseFly eBee                                                                                                                                                                                  & 3                         & {\cite{s17092007}}, {\cite{10.1080/01431161.2016.1252475}}, {\cite{jimaging4110132}}                                                 \\
                                                                                                                                                                                               &                           &                                                                             \\
MD4-1000                                                                                                                                                                                       & 3                         & {\cite{rs10020285}}, {\cite{7849987}}, {\cite{PEREZORTIZ2015533}}                                                 \\
                                                                                                                                                                                               &                           &                                                                             \\
Micro Aerial Vehicle                                                                                                                                                                           & 1                         & {\cite{10036378}}                                                                     \\
                                                                                                                                                                                               &                           &                                                                             \\
MikroKopter JR11X                                                                                                                                                                              & 1                         & {\cite{agronomy11071435}}                                                                     \\
                                                                                                                                                                                               &                           &                                                                             \\
CarbonCore Cortex X8 Octocopter                                                                                                                                                                & 1                         & {\cite{land10010029}}                                                                    \\
                                                                                                                                                                                               &                           &                                                                             \\
ATI AgBot                                                                                                                                                                                      & 1                         & {\cite{agronomy11101909}}                                                                    \\ \hline
\label{tab3}
\end{tabular}
\end{table}

The RedEdge-M Camera is used in 5 of the 21 studies. It is a professional multispectral sensor that weighs approximately 173 grams and captures blue, green, red, red edge, near-IR (global shutter, narrowband) spectral bands. The Sequoia sensor, Canon S110 NIR, and TetraCam mini MCA-6 were also used prominently in the studies. The features and references to these cameras are illustrated in Table \ref{tab4}. 

\begin{table}[ht]
\caption{Camera Models used in the selected research papers}
\begin{tabular}{@{}ccccccc@{}}
\toprule
\textbf{Camera}                                                             & \textbf{\begin{tabular}[c]{@{}c@{}}Weight\\ (g)\end{tabular}} & \textbf{\begin{tabular}[c]{@{}c@{}}Dimensions\\ (mm)\end{tabular}} & \textbf{Power}                                                                                                & \textbf{Spectral Bands}                                                                                                                                                                                                                                                   & \textbf{\begin{tabular}[c]{@{}c@{}}Ground \\ Sample \\ Distance \\ \\ (GSD)\end{tabular}} & \textbf{Reference}                                                                            \\ \midrule
RedEdge-M                                                                   & 170                                                           & 94x63x46                                                           & \begin{tabular}[c]{@{}c@{}}4.2-15.8 V, \\ 4W Nominal, \\ 8W Peak\end{tabular}                                 & \begin{tabular}[c]{@{}c@{}}R, G, B, \\ Red Edge, \\ NIR\end{tabular}                                                                                                                                                                                      & 8 cm at 120 m                                                                             & \begin{tabular}[c]{@{}c@{}}{\cite{rs14174197}}, {\cite{9839758}}, \\ {\cite{rs10091423}}, {\cite{SU2022106621}},\\  {\cite{Wang2022WeedDE}}\end{tabular} \\\\
\begin{tabular}[c]{@{}c@{}}Sequoia \\ Sensor\end{tabular}                   & 72                                                            & 59x41x28                                                           & \begin{tabular}[c]{@{}c@{}}5-6V, \\ Minimal threshold\\  12W,\\ Optimal \\ recommendation \\ 15W\end{tabular} & \begin{tabular}[c]{@{}c@{}}R, G, \\ Red Edge, \\ NIR\end{tabular}                                                                                                                                                                                               & 1.9 cm at 70 m                                                                            & \begin{tabular}[c]{@{}c@{}}{\cite{10036378}}, {\cite{agronomy13030830}},\\ {\cite{8936091}}, {\cite{rs10091423}}\end{tabular}  \\               \\
\begin{tabular}[c]{@{}c@{}}Canon \\ S110 NIR\end{tabular}                   & 198                                                           & 99x59x27                                                           &  N/M                                                                                                         & \begin{tabular}[c]{@{}c@{}}R, G, NIR\end{tabular}                                                                                                                                                                                                            & 3.5 cm at 100 m                                                                           & \begin{tabular}[c]{@{}c@{}}{\cite{s17092007}}, {\cite{10.1080/01431161.2016.1252475}}, \\ {\cite{jimaging4110132}}\end{tabular}   \\                     \\
\begin{tabular}[c]{@{}c@{}}TetraCam \\ mini \\ MCA-6\end{tabular}           & 700                                                           & 131.4x78.3x87.6                                                    & 5.4 W                                                                                                         &  R, G, B, NIR                                                                                                                                                                                                                                                                          &  5 cm at 100 m                                                                                         & \begin{tabular}[c]{@{}c@{}}{\cite{agronomy11071435}}, {\cite{7849987}}, \\ {\cite{PEREZORTIZ2015533}}, {\cite{rs71012793}}\end{tabular}  \\                      \\
\begin{tabular}[c]{@{}c@{}}Micasense \\ Altum \\ Multispectral\end{tabular} & 49                                                            & 63.5x46x14.03                                                      & 10 W                                                                                                          & \begin{tabular}[c]{@{}c@{}}R, G, B, \\ Red Edge, \\ NIR, Thermal\end{tabular}                                                                                                                                                                          & 5.2 cm at 120 m                                                                           & {\cite{rs15061633}}, {\cite{agronomy11101909}}\\                                                                             \\
\begin{tabular}[c]{@{}c@{}}Sony\\ ILCE-6000\end{tabular}                    & 413                                                           & 120x66.9x45.1                                                      & N/M                                                                                                              & \begin{tabular}[c]{@{}c@{}}R, G, NIR\end{tabular} & 0.35 cm at 22 m                                                                                           & {\cite{rs10020285}}   \\                                                                                    \\
\begin{tabular}[c]{@{}c@{}}Cubert \\ UHD-185\end{tabular}                   & 470                                                           & 28x6.5x7                                                           & DC 12V, 15W                                                                                                   & 450 - 950 nm                                                                                                                                                                                                                                                              &  4.8 cm at 120 m                                                                                         & \cite{land10010029}{\cite{DENG2018298}}                                                   \\                                   \\
\begin{tabular}[c]{@{}c@{}}Resonon \\ Pika L\end{tabular}                   & 600                                                           & 115x104x66                                                         & 3.4 W                                                                                                         & 400 - 1000 nm                                                                                                                                                                                                                                                             & 3.7 cm at 50 m                                                                                          & {\cite{Scherrer2019HyperspectralIA}, \cite{online4}}            \\                                                                          \\
\begin{tabular}[c]{@{}c@{}}Agrocam \\ NDVI\end{tabular}                     & 78                                                            & 59x41.5x36                                                         & N/M                                                                                                              &                                                                                                                                                                                                                                                                           NIR, G, B&  0.33 cm at 4.5 m                                                                                          & {\cite{DENG2018298},  \cite{9787455}}                                                                                      \\ \bottomrule
\label{tab4}
\end{tabular}
\end{table}

\section{RQ2: What features are extracted from the captured data}
The features extracted from the collected data are one of the most important aspects that must be analyzed in order to train efficient models, capable of accurately identifying weeds in the field.

Normalized Difference Vegetation Index (NDVI), Green Normalized Difference Vegetation Index (GNDVI), Normalized Difference Red Edge Index (NDRE), Green Coverage Index (GCI), Modified Soil-Adjusted Vegetation Index (MSAVI), and Excess Green Vegetation Index (ExG) were vegetation indices which were prominently used in several studies.  
These indicators are adopted to collect vital information about the crop's health, chlorophyll content, water level, weed detection, and other parameters of research interest using different sensors.

NDVI is a greenness index that is derived from computing the ratio between the TOA reflectance of a red band, and the near-infrared (NIR) band \cite{voor19}. The underlying concept of this index is that healthy and green crops that have a high amount of chlorophyll tend to absorb red light while reflecting NIR during photosynthesis, and vice-versa. This relationship between light reflectance and chlorophyll content is used to get insight into the crop's distribution, health, and disease spread across a field, etc. GNDVI, NDRE, GCI, and SAVI are modified versions of NDVI that also employ red-edge or NIR bands. ExG is an RGB vegetation index that uses RGB chromatic coordinates to distinguish between plants, and non-plant backgrounds such as weeds, soil, etc. \cite{Woebbecke1994ColorIF}. 

Table \ref{tab5} illustrates the basic formulae and references of these vegetation indices \cite{rs11131548}, where \textit{N} is the Near-infrared band reflectance, \textit{R} is the red band reflectance, \textit{G} is the green band reflectance, \textit{E} is the Red Edge band reflectance, and \textit{B} is the blue band reflectance.  

\begin{table}[H]
\caption{Formulae and References of the features used in studies}
\begin{tabular}{@{}c|ccc@{}}
\toprule
\textbf{Type} & \multicolumn{1}{c}{\textbf{Name}}                                        & \textbf{Equation}                                                                     & \multicolumn{1}{l}{\textbf{Reference}} \\ \midrule
NIR           &  Normalized Difference Vegetation Index (NDVI)                            & (N-R) / (N+R)                                                                     & {\cite{online4}}                               \\
              & \multicolumn{1}{l}{Green Normalized Difference Vegetation Index (GNDVI)} & (N-G) / (N+G)                                                                     & {\cite{DENG2018298}}                               \\
              & Normalized Difference Red Edge Index (NDRE)                              & (N-E) / (N+E)                                                                     & {\cite{9787455}}                               \\
              & Green Chlorophyll Index (GCI)                                            & (N / G) -1                                                                           & {\cite{rs71012793}}                               \\
              & \multicolumn{1}{l}{Modified Soil Adjusted Vegetation Index (MSAVI)}      & \multicolumn{1}{l}{$2N+1-\sqrt{(2N+1)^{2}-8(N-R)}/2$}& {\cite{msavi}}                               \\ & \\
              \hline
RGB           & ExG (Excess Green)                                                       & 2G-R-B                                                                         & {\cite{HAMUDA2016184}}                               \\ \bottomrule
\label{tab5}
\end{tabular}
\end{table}

Apart from these, texture, brightness, and plant height were also used in some studies as additional features to obtain better results \cite{s17092007, rs10020285, jimaging4110132}, as shown in Table \ref{tab6}.
Texture has been shown to provide enhanced results in object-based image analysis, with entropy texture measure giving the highest scores in decision trees \cite{4773214}. The use of features like spatial texture and estimated vegetation height to map milk thistle (Silybum marianum) weed patches using UAVs has also shown improvements in overall classification accuracy. Since texture is a relatively simpler and easy-to-compute feature compared to plant height, it may be employed in future studies related to weed classification \cite{jimaging4110132}.

\begin{table}[H]
\caption{\label{tab:table1} Features used in the selected research papers for training machine learning models}
\begin{tabular}{c|l}
\hline
\textbf{Research Paper}                & \multicolumn{1}{c}{\textbf{Features Used}}                                                                                                                                                                                                                                                         \\ \hline
{\cite{s17092007}}                                & 3 Spectral bands with texture                                                                                                                                                                                                                                                                      \newline                                                                                                                                                                                                                                                                                                  \\
{\cite{rs15061633}}                                & 5 Reflectance bands and 6 vegetation indices (NDVI, GNDVI, NDRE, GCI, MSAVI, ExG)                                                                                                                                                                                                                  \newline                                                                                                                                                                                                                                                                                               \\
{\cite{10036378}}                                & NIR and Red channel images were used to generate NDVI images                                                                                                                                                                                                                                       \\
                                       &                                                                                                                                                                                                                                                                                                    \\
{\cite{rs10020285}}                                & \begin{tabular}[c]{@{}l@{}}NIR/G, Mean Red, Mean Green, Mean NIR, Brightness, NIR standard deviation, and CHM \\ standard deviation values were extracted for every object making up the training set.\end{tabular}                                                                                \\
                                       &                                                                                                                                                                                                                                                                                                    \\
{\cite{Scherrer2019HyperspectralIA}}                               & \begin{tabular}[c]{@{}l@{}}- 19,188,642 spectra were extracted from 67 hyperspectral images\\ - The imager provided a series of images with two spatial dimensions and one spectral dimension.\\  Each spectrum was then extracted from its corresponding pixel in the spatial plane.\end{tabular}                                                                                                                                                                                                                                                                                                \\
                                       &                                                                                                                                    \\
{\cite{7849987}}, {\cite{PEREZORTIZ2015533}}, {\cite{agronomy11101909}}, {\cite{SU2022106621}} & Blue (475/20 nm), Green (560/20 nm), Red (668/10 nm), RedEdge (717/10 nm) and NIR (840/40 nm)                                                                                                                                                                                                      \newline                                                                                                                                                                                                                                                                                                   \\
{\cite{Wang2022WeedDE}}                               & RGB + 5-band multispectral Images                                                                                                                                                                                                                                                                 \newline                                                                                                                                                                                                                                                                                                \\
{\cite{jimaging4110132}}                               & G, R, NIR, Texture, Plant Height                                                                                                                                                                                                                                                                   \\ \hline
\label{tab6}
\end{tabular}
\end{table}

\section{RQ3: Which machine learning algorithms have been applied for weed detection?}

The machine learning algorithms used in the research papers have been divided into 7 categories: Neural Networks, Support Vector Machines, Ensemble Learning, Clustering, Bayesian Models, Instant Based Models, and Others. This classification has been made based on other related literature reviews \cite{s23073670}. The Algorithm Categories, Applied Algorithms, and the number of times the algorithms have been used are shown in Table \ref{tab7}. The number of times the algorithms have been used is greater than the total number of papers because some papers make use of more than one machine learning algorithm in order to perform comparative analyses among them.

One of the studies utilised a semi-supervised generative adversarial network (GAN) on the weedNet dataset consisting of multispectral crop and weed images collected by a micro aerial vehicle (MAV), which enables pixel-level classification of all the multispectral images obtained \cite{8936091}. The method employed a generator network that produces realistic images as additional training data, which was then used to train a multi-class classifier that acted as a discriminator. This classifier was tested on the weedNet dataset, and the outcome showed that the overall approach enhances the accuracy of the classification results.

\begin{table}[H]
\caption{Machine learning algorithms used in the selected research papers}
\begin{tabular}{lllll}
\cline{1-3}
\textbf{Algorithms}  & \textbf{Applied Algorithms}                                                                                                                                                                                                                                                                                                             & \textbf{\# of Times Used} &  &  \\ \cline{1-3}
Neural Networks      & \begin{tabular}[c]{@{}l@{}}Auto-encoder, One-Class Self-Organizing Map, Hierarchical \\ Neural Networks (HNN), General Adversarial Networks (GAN), \\ Convolutional Neural Networks (CNN),  Artificial Neural \\ Networks (ANN), Deep Neural Networks (DNN), \\ Feed-Forward Neural Networks (FFNN), Patch-Image-Based CNN\end{tabular} & \multicolumn{1}{c}{13}    &  &  \\
                     &                                                                                                                                                                                                                                                                                                                                         &                           &  &  \\
SVMs                 & \begin{tabular}[c]{@{}l@{}}SVM, One-Class SVM, Semi-supervised SVM, Linear SVM, \\Kernel SVM, HOG + SVM, Local Binary Pattern (LBP) SVM\end{tabular}                                                                                                                                                                                    & \multicolumn{1}{c}{13}    &  &  \\
                     &                                                                                                                                                                                                                                                                                                                                         &                           &  &  \\
Ensemble Learning    & Random Forest, XGBoost                                                                                                                                                                                                                                                                                                                  & \multicolumn{1}{c}{4}     &  &  \\
                     &                                                                                                                                                                                                                                                                                                                                         &                           &  &  \\
Clustering           & K-means, Repeated K-means                                                                                                                                                                                                                                                                                                               & \multicolumn{1}{c}{4}     &  &  \\
                     &                                                                                                                                                                                                                                                                                                                                         &                           &  &  \\
Bayesian Model       & Maximum Likelihood Classification                                                                                                                                                                                                                                                                                                       & \multicolumn{1}{c}{3}     &  &  \\
                     &                                                                                                                                                                                                                                                                                                                                         &                           &  &  \\
Instant Based Models & K-nearest Neighbor                                                                                                                                                                                                                                                                                                                      & \multicolumn{1}{c}{3}     &  &  \\
                     &                                                                                                                                                                                                                                                                                                                                         &                           &  &  \\
Others               & \begin{tabular}[c]{@{}l@{}}OBIA, Step-wise Linear Discriminant Analysis, \\ One-Class Principal Component Analysis\end{tabular}                                                                                                                                                                                                         & \multicolumn{1}{c}{5}     &  &  \\ \cline{1-3}
\label{tab7}
\end{tabular}
\end{table}

Another paper proposed a methodology that involved using patch images of a sesame field for classification, which was carried out through a newly developed convolutional neural network (CNN) called SesameWeedNet \cite{DENG2018298}. This CNN model had 6 convolutional layers and was specifically designed to work efficiently and effectively with smaller patch images. The system filtered out patches that contained only background, as the vegetation was already detected in the initial step. This approach effectively simplified the classification task from a 3-class (background, crop, and weed) problem to a 2-class (crop and weed) problem. It achieved an accuracy of 97\% on a sesame field in Pakistan, using Agrocam NDVI camera and Phantom 3 UAV.

A study that was conducted on three crops, maize, peanut seedlings, and wheat, proposed improved transfer neural networks optimised using two bionic optimization algorithms, i.e., (1) Particle Swarm Optimization and (2) Bat Algorithm, respectively \cite{Wang2022WeedDE}. Both of these neural networks were compared with a self-constructed CNN based on model-agnostic meta-learning (MAML) and histogram of oriented gradient + support vector machine (HOG + SVM). The results showed that the combination of learning rates through the Bat Algorithm turned out to be the best among the three, attaining 99.53\% accuracy for the multispectral images. 

According to the results of the analysis of all the papers, the algorithm categories being used the most are Neural Networks and Support Vector Machines, which leads to the conclusion that these two algorithms are more effective when it comes to weed detection using multispectral UAV imagery and machine learning.

\section{RQ4: Which are the evaluation approaches and parameters used?}

Evaluation parameters or evaluation metrics are essential to gauge exactly how well the model is generalizing over unknown data. There are various metrics that can be used based on the type of data and machine learning model used in the study. Accuracy, which is the proportion of correctly classified instances, provides a broad view of overall correctness. The F1 score combines precision and recall, offering a balanced measure, especially vital when dealing with imbalanced classes. Area Under the Curve (AUC) quantifies a binary classifier's ability to distinguish between positive and negative classes. User's and Producer's accuracies focus on correctly classified instances within confusion matrices. The Kappa statistic gauges agreement while considering chance in classifications. Precision emphasises accurate positive predictions, while recall measures a model's ability to identify all positives. Regression metrics like Mean Absolute Error (MAE) and Mean Squared Error (MSE) gauge prediction accuracy and error. The KHAT Statistic evaluates classifier agreement in specialised contexts. Intersection over Union (IoU) finds use in object detection by calculating the region overlap. Lastly, mean entropy assesses uncertainty in probability distributions. Each metric contributes uniquely to understanding and evaluating diverse aspects of machine learning models.

Multiple evaluation metrics like Global Accuracy, Kappa Index, F1 Score, and AUC have been used for the evaluation of a GAN to detect weed patches in a sugar beets field \cite{8936091}. MAE has been used to evaluate the performance of multiple machine learning models like k-means, Rk-means, SVM, SS-SVM, LinSVM, and K-Nearest Neighbours (KNN) to classify weeds in a field of sunflowers \cite{7849987, PEREZORTIZ2015533}. Some other metrics like user's accuracy and producer's accuracy have been used to identify weeds in a field of sorghum using object-based image analysis (OBIA) and step-wise linear discriminant analysis \cite{agronomy11071435}, and to detect S.Marianum weed in a field with random vegetation using the Maximum Likelihood algorithm \cite{10.1080/01431161.2016.1252475, jimaging4110132} and the Minimum Distance algorithm \cite{jimaging4110132}.

\begin{table}[ht]
\caption{Evaluation parameters used in the selected research papers }
\begin{tabular}{ccl}
\hline
\textbf{Evaluation Parameters} & \multicolumn{1}{c}{\textbf{Number of Times Used}} & \multicolumn{1}{c}{\textbf{References}}                                                                                                       \\ \hline \\
Accuracy                       & 15                                                & \makecell[l]{\{\cite{s17092007}}, {\cite{rs15061633}}, {\cite{10036378}}, {\cite{agronomy13030830}}, {\cite{agronomy11071435}}, {\cite{rs10020285}}, {\cite{10.1007/978-3-030-29888-3_45}}, {\cite{land10010029}}, {\cite{agronomy11101909}}, {\cite{Scherrer2019HyperspectralIA}}, {\cite{DENG2018298}}, {\cite{SU2022106621}}, \\{\cite{10.1080/01431161.2016.1252475}}, {\cite{Wang2022WeedDE}}, {\cite{jimaging4110132}} \\
                               &                                                   &                                                                                                                                               \\
F1                             & 3                                                 & {\cite{rs15061633}}, {\cite{agronomy13030830}}, {\cite{8936091}}                                                                                                                     \\
                               &                                                   &                                                                                                                                               \\
AUC                            & 3                                                 & {\cite{agronomy13030830}}, {\cite{rs10091423}}, {\cite{DENG2018298}}                                                                                                                   \\
                               &                                                   &                                                                                                                                               \\
User's Accuracy                & 3                                                 & {\cite{agronomy11071435}}, {\cite{10.1080/01431161.2016.1252475}}, {\cite{jimaging4110132}}                                                                                                                   \\
                               &                                                   &                                                                                                                                               \\
Producer's Accuracy            & 3                                                 & {\cite{agronomy11071435}}, {\cite{10.1080/01431161.2016.1252475}}, {\cite{jimaging4110132}}                                                                                                                   \\
                               &                                                   &                                                                                                                                               \\
Kappa Statistic                & 3                                                 & {\cite{agronomy13030830}}, {\cite{10.1080/01431161.2016.1252475}}, {\cite{jimaging4110132}}                                                                                                                   \\
                               &                                                   &                                                                                                                                               \\
Precision                      & 2                                                 & {\cite{rs15061633}}, {\cite{SU2022106621}}                                                                                                                             \\
                               &                                                   &                                                                                                                                               \\
Mean Absolute Error            & 2                                                 & {\cite{7849987}}, {\cite{PEREZORTIZ2015533}}                                                                                                                            \\
                               &                                                   &                                                                                                                                               \\
Recall                         & 2                                                 & {\cite{rs15061633}}, {\cite{SU2022106621}}                                                                                                                             \\
                               &                                                   &                                                                                                                                               \\
Mean Square Error              & 1                                                 & {\cite{10036378}}                                                                                                                                       \\
                               &                                                   &                                                                                                                                               \\
KHAT Statistic                 & 1                                                 & {\cite{agronomy11071435}}                                                                                                                                       \\
                               &                                                   &                                                                                                                                               \\
Intersection over Union (IoU)  & 1                                                 & {\cite{9839758}}                                                                                                                                      \\
                               &                                                   &                                                                                                                                               \\
Mean Entropy                   & 1                                                 & {\cite{9839758}}                                                                                                                                      \\ \hline
\label{tab8}
\end{tabular}
\end{table}

In the selected research papers, a comprehensive set of 13 distinct evaluation metrics has been used. This collection of metrics has been invoked in a total of 40 instances, as detailed in Table \ref{tab8}. The most used evaluation metric was Accuracy, which was used 15 times, i.e., for almost 38 \% of the total evaluations. On the other hand, the least used metrics were Mean Square Error (MSE), the KHAT Statistic, Intersection-over-Union, and Mean Entropy, which were just used once each.

The majority of the conducted studies are engaged in binary classification tasks that aim to distinguish between crops and weeds. In the context of classification problems, the adoption of MSE as an evaluation metric is not commonly favored. This disinclination is rooted in the inherent assumption of MSE that the underlying data adheres to a normal distribution, a premise that does not hold universally true in practice. This discernible trend is mirrored in the selected research papers, wherein only a solitary study employs MSE for evaluation purposes \cite{10036378}. Furthermore, it is noteworthy that even in this singular instance, MSE is not utilised in isolation, but rather in conjunction with the accuracy metric.

\section{RQ5: What are the major challenges in this domain?}

One of the major challenges in this field of study is the unavailability of a large quantity and variety of training data. It was observed that the general machine learning algorithms like XGBoost, SVM, KNN, and Random Forest did not work as well for different terrains and different types of hawkweeds \cite{rs15061633}. The unavailability of high-resolution images for accurate labeling, and the time and expertise required to label images accurately is, therefore, a major hindrance. Moreover, the color and texture similarity of neighboring species presents challenges for accurate labeling of the images.
In addition to this, for algorithms like OBIA \cite{rs10020285}, there isn't much research available that can provide accurate spatial measurement methods, and the conventional accuracy assessment techniques, which rely on site-specific reference data, cannot offer this type of information.

Another challenge that is faced is with respect to multispectral imaging, which is that the spectral properties of certain crops or weeds may turn out to be quite similar to each other, which may lead to improper training of the machine learning model. Such a challenge was faced in a study where the autoencoder and OC-PCA algorithms overestimated the S. marianum weeds in the northern region of the study area, which was actually dominated by A. sterilis \cite{s17092007}. This error might have occurred because the A. sterilis plants managed to stay healthy during the late spring season while the rest of the vegetation in the area was decaying. This led to the spectral signatures of the two species to become similar to each other, thereby leading to confusion. 

In addition to the above, the elimination of cluttered backgrounds from multispectral images is also a very important step in order to extract only useful information \cite{10036378}. The selection of appropriate kernel size to enhance the contrast of only salient crops is another hurdle in the way of obtaining good accuracy.

Furthermore, a big challenge pertaining to UAVs is the height of the flight, which is a very important parameter to be considered. Flying the UAV too high can result in low-resolution and unclear images, which would not be of any use in training the algorithm \cite{7849987}. However, if the UAV is flown too low, it can lead to problems like an increase in collision risk with trees or other obstructions, limited coverage area, and a decrease in flight time because of the increase in power consumption that happens when flying at a low altitude.

Another big challenge when it comes to UAVs is that high-resolution aerial images to investigate weeds can only be acquired in areas that are not covered by the canopy of trees \cite{land10010029}. Furthermore, UAV devices capable of producing images with sufficiently high spatial resolution have limitations regarding coverage. Therefore, research has to be confined to a local scale.
All the aforementioned challenges are given in a concise manner in Table \ref{tab9}.

\begin{table}[ht]
\caption{Challenges in the field of study as identified from the selected research papers}
\begin{tabular}{c|l}
\hline
\textbf{S.No.} & \multicolumn{1}{c}{\textbf{Challenges}}                                                      \\ \hline
\\
1.             & Unavailability of a large volume of training data                                           \\
               &                                                                                             \\
2.             & Unavailability of a variety of data (with different weather conditions, terrains, etc.)      \\
               &                                                                                             \\
3.             & Similar spectral signatures of crops and weeds, or among different types of weeds           \\
               &                                                                                             \\
4.             & Cluttered backgrounds in multispectral images                                               \\
               &                                                                                             \\
5.             & Selection of appropriate kernel size for images                                             \\
               &                                                                                             \\
6.             & Determining optimum heights of UAV flights                                                  \\
               &                                                                                             \\
7.             & Difficult to capture UAV images if crops are covered by canopies or other such obstructions \\
               &                                                                                             \\
8.             & High-resolution imaging UAVs usually have less coverage                                     \\ \hline

\label{tab9}
\end{tabular}
\end{table}

\section{Discussion}

The systematic review has been conducted in accordance with the PRISMA guidelines, with an iteratively formed search strategy. Research databases were selected on the basis of previously conducted studies that analyzed the most prominent sources for UAV-based agricultural studies, and a set of predefined inclusion and exclusion criteria to identify the specific and most relevant articles for meta-analysis. The entire search procedure has been discussed in detail, and all associated references and data have been included to ensure replicability and reliability. 

The study was initiated with the objective of analyzing the potential of identifying weeds in agricultural fields, specifically with the help of machine-learning-based interventions, and UAV-acquired multispectral and hyperspectral images. The study provides a comprehensive view of the various aspects involved in the process which includes the most effective machine learning models, features, evaluation parameters, and the devices used. Moreover, it simultaneously highlights relevant work and implementations from the selected studies. 

The study shows an upward trend in the number of researchers who have employed hyperspectral and multispectral data to predict weeds in agricultural fields in the past years. 

For the sake of objectivity, studies that only employed RGB datasets were excluded from the analysis, however, in the future, studies could be conducted with the objective of comparing different approaches (individual and hybrid) to capture data and their respective efficiency. 

Computational expense and resource requirement analysis for different methods of data acquisition and modeling could also be explored, such as the resource-friendly model ProtoNN \cite{inproceedings1}. 

It is important to note that there are some limitations to the scope of this review. The first limitation lies in the fact that studies of only the English language were chosen for this review, and others were not taken into consideration. This may have led to the review being devoid of the results of some relevant studies published in different languages. Secondly, the review may have been limited by the search string and the three sources selected for obtaining the research papers, even though the study tries to minimise this limitation by selecting databases that have been found to be most relevant to the topic of interest, and technologies that have been studied.  

\section{Conclusion and Recommendations}
\subsection{\emph{Conclusions}}

A systematic review was conducted on the technologies that combine multispectral data acquired with the help of Unmanned Aerial Vehicles (UAVs) and Machine Learning techniques for weed detection in agricultural fields. After the title and abstract screening and the full-text review based on the selection criteria specified, a total of 21 studies were chosen to be used in the review. These papers were then used to answer the 5 review questions as given in Section V(A).

The RedEdge-M Camera was the most used imaging device and was used in around 24\% of the studies. Different models of the DJI UAVs, such as DJI Inspire 2, DJI Mavic, DJI Matrice 600, DJI Matrice 100, and DJI S1000, were the most used  UAVs and were used for image collection in approximately 47\% of all the studies. The combination of features extracted from the captured images the most number of times is Blue (475/20 nm), Green (560/20 nm), Red (668/10 nm), RedEdge (717/10 nm) and NIR (840/40 nm), which was used in 19\% of the studies. The machine learning algorithms most applied for weed detection turned out to be Neural Networks and Support Vector Machines, each of which was used in about 29\% of the studies considered. The most popular model evaluation metric is accuracy, which is being used to evaluate 40\% of all the machine learning models in the chosen studies. The main challenges as stated in the research studies were the unavailability of a large volume and variety of training data, confusion due to similar spectral signatures of crops and weeds or among different types of weeds, decluttering of background in multispectral images, selection of appropriate kernel size for images, determining the optimum height for the UAV flights, difficulty in capturing aerial images due to obstructions like canopies over the crops and the problem of low-coverage in high-resolution imaging UAVs.
According to the aforementioned conclusions, some recommendations are listed in the following section.

\subsection{\emph{Recommendations}}

The significant recommendations to carry out similar systematic reviews include relaxing the search string to include more studies. One can also widen the scope of the search by including more databases to obtain research studies. The search can further be broadened to include more books, reports, conference papers, proceedings, ongoing studies, and so on. Specific to this review, studies with solely RGB data were excluded, hence further reviews could include those studies as well. For higher-level reviews, other techniques to carry out analysis can also be explored.

Future research endeavors within this domain may be channeled towards the development and evaluation of real-time weed detection systems, capable of providing instantaneous feedback to the farmers. Furthermore, the integration of Internet of Things (IoT) technologies into these systems also holds the potential of establishing a seamless infrastructure for data collection, transmission, and analysis. Research efforts can also be directed towards the optimization of resource-efficient algorithms in IoT devices, further expediting the decision-making processes.

This holistic integration of real-time weed detection and IoT technologies could also facilitate the creation of real-time weedicide-spraying systems. These systems utilise target spraying, a method characterised by the precise application of weedicides exclusively to the areas where weeds are detected. Target spraying offers several advantages, including cost savings for farmers, mitigation of herbicide resistance in weeds, and a reduction in the ecological footprint associated with herbicide usage.

Our study selects papers that conducted on-field experiments for data collection, and the trained models that are used show a promising potential to detect weeds, with high accuracy (using appropriate ML models) and a broader range of view (using UAVs to gather images). However, it is essential to perform more experiments and reviews that study the feasibility and scalability of implementing such systems over a wider, more diverse, and unpredictable field of study. The systems should be tested across multiple locations, with different geographical, agricultural, and economic parameters to gauge the overall efficiency of the systems. 

Moreover, it is recommended to conduct more research and field experiments that investigate the long-term economic, and environmental impact of employing such systems as a replacement to conventional weed management techniques. Systems could be tested for different crops and weeds at different locations to identify the actual result on the crop-to-weed ratio, weed density and diversity, crop yield, yield quality, and its effect on environmental and human health. 

Data unavailability is another key challenge that must be addressed to utilise a more diverse set of ML and Deep Learning (DL) algorithms, efficiently.
Curating and sharing well-labeled, high-quality, and standardised datasets would be extremely beneficial in validating and benchmarking novel solution strategies in this domain. 

Additionally, there exists a scope for the exploration of effective ML and DL algorithms. Comparative studies can be conducted with the aim of assessing the accuracy of diverse learning models and their associated parameters when applied to identical datasets. Such comparative investigations would serve the purpose of determining the most suitable models for specific use cases. 

\section{Disclosure of Interest}
The authors report no conflict of interest. 

\printbibliography

\end{document}